\documentclass[aps,prl,twocolumn,amsmath,amssymb]{revtex4}
\usepackage{times}
\usepackage{latexsym}
\usepackage{graphicx}
\usepackage{amsmath,amssymb}
\usepackage{verbatim,bbm}
\usepackage{hyperref}

\begin{document}

\title{Parameter estimation with almost no public communication\\
for continuous-variable quantum key distribution}
\author{Cosmo Lupo, Carlo Ottaviani, Panagiotis Papanastasiou, Stefano Pirandola}
\affiliation{Department of Computer Science, University of York, York YO10 5GH, UK}

\begin{abstract}
One crucial step in any quantum key distribution (QKD) scheme is
parameter estimation. In a typical QKD protocol the users have to sacrifice 
part of their raw data to estimate the parameters of the communication 
channel as, for example, the error rate.
This introduces a tradeoff between the secret key rate and the accuracy
of parameter estimation in the finite-size regime.
Here we show that continuous-variable (CV) QKD is not subject to
this constraint as the whole raw keys can be used for both parameter
estimation and secret key generation, without compromising the security.
First we show that this property holds for measurement-device
independent (MDI) protocols, as a consequence of the fact that in an 
MDI protocol the correlations between Alice and Bob are post-selected 
by the measurement performed by an untrusted relay.
This result is then extended beyond the MDI framework by exploiting the
fact that MDI protocols can simulate device-dependent one-way QKD with
arbitrarily high precision.
\end{abstract}

\maketitle

{\it Introduction:--}
Quantum key distribution (QKD) exploits quantum physics
to distribute secret keys between distant users 
that have access to an insecure quantum communication channel \cite{Gisin,QKD-RMP,RMP,QKDCV}. 
These secret keys can then be used as one-time pads to achieve 
information-theoretically secure communication \cite{OTP}.
A QKD protocol is an explicit recipe to achieve this goal and 
typically comprises two parts: 
a quantum part where quantum signals are transmitted through a 
quantum channel connecting two authenticated users (typically
named Alice and Bob) and then measured at the output of the channel; 
a classical part where local classical information about the state 
preparation and measurement outputs are processed to extract a common, 
secret key.

One crucial part of classical post-processing is parameter estimation,
a routine aiming at obtaining information about the quantum channel
connecting Alice to Bob. The task of parameter estimation is similar
to quantum channel (or state) tomography (see e.g.\ Ref.\ \cite{tomog} and
references therein), though in this case one 
is not interested in obtaining a full description of the quantum channel, 
but only in those features that are relevant for the security of the QKD protocol.
Once the quantum channel is estimated, the principles
of quantum mechanics impose an upper bound on the amount of information that
has possibly leaked to a potential eavesdropper.
In general, local information without classical communication
is not sufficient to perform neither parameter estimation nor quantum 
state tomography \cite{Laflamme,Zeng,Linden}. 
For this reason, it is required that Alice and Bob exchange part 
of their local data in order to perform parameter estimation.
Obviously, all the classical data that are communicated through an insecure 
channel must be considered compromised. 
It follows that the more data are used for parameter estimation, the 
lower is the final secret key rate. Viceversa, if less data are used for
parameter estimation, then statistical errors will make the estimation less
accurate.

In this Letter we show that for continuous-variable (CV) QKD protocols
(as for example those in Refs.\ \cite{Lev1,Lev2,Chris,RCI,Raul,carlo2017,Furrer,Lev2013,thermal,NLA,Lev2012})
one can use, without loss of security, the whole local data for both parameter estimation
and secret key extraction.
This result is a consequence of a characteristic features of CV QKD:
that the knowledge of the covariance matrix (CM) of the field quadratures
is in general sufficient to assess the security of a CV QKD protocol \cite{Wolf,Raul}.
To prove this result we consider the framework of measurement-device
independent (MDI) QKD, first introduced to achieve security against
side-channel attacks on the measurement devices \cite{Curty,Sam}.
Then, the result is extended to one-way CV QKD protocols by exploiting
the fact that the latter can be simulated by an MDI protocol up to an
arbitrarily small error \cite{carlo}. 

In previous works, other authors have discussed a way to
use the whole raw keys for both parameter estimation and secret key
extraction. This can be achieved if the users first obtain
a rough estimate of the error rate (or of the signal-to-noise ratio) and 
then exploit it to perform error correction {\it before} parameter estimation \cite{Pacher,Lev1}.
Our approach is independent and conceptually different as we do not 
need a rough estimate of the channel parameters and we do not rely
on doing error correction before parameter estimation.


{\it The structure of a QKD protocol:--}
Up to a few conceptually significant advancements, the structure of QKD protocols 
has remained mostly constant since the first QKD protocol 
was proposed by Bennett and Brassard in 1984 (BB84) \cite{BB84}.
A typical QKD protocol consists of seven basic operations:
(1) State preparation:
Alice generates a sequence of $n$ symbols, for each symbol she 
prepares a suitable quantum codeword. For example, in the original
BB84 protocol Alice encodes a bit value $X \in \{ 0, 1\}$ in one
qubit either using the computational basis $\{ |0\rangle, |1\rangle\}$
or the diagonal basis $\{ |+\rangle, |-\rangle\}$.
(2) Communication:
The quantum states are transmitted through an insecure quantum 
communication channel.
(3) Measurement:
Bob measures the quantum states coming out of the communication 
channel. For example, in the BB84 protocol Bob obtains a bit
value $Y \in \{0,1\}$ by either measuring in the computational or 
diagonal basis.
(4) Sifting:
For each signal transmitted, Alice and Bob publicly announce whether 
they have employed the computational or diagonal basis. 
Then they only retain the data 
corresponding to matching choices for preparation and measurement.
The sifted data represent the local raw keys of Alice and Bob.
(5) Parameter estimation:
Alice and Bob publicly agree on a
subset of their local data to estimate the parameters of the channel.
For example, Bob sends to Alice
a fraction $f$ of his data, so that she can estimate the probability
of error.
Obviously, all the data sent through the public channel for 
parameter estimation are compromised and cannot be used
for secret key extraction: the final rate will thus be reduced by a 
factor $1-f$.
(6) Error correction:
Alice sends to Bob
error-correcting information. Bob can combine this information 
with his local data to reconstruct Alice's raw keys up to a small
error (direct reconciliation).
(7) Privacy amplification:
Alice and Bob apply a hash function to obtain a shorter key which
a potential eavesdropper has virtually no information about.

During the three decades that separate us from BB84, 
several main conceptual development of QKD has been introduced.
One of the main advancements in QKD has been the introduction of CV 
protocols \cite{Hillery,Cerf}, in which information is encoded in continuous degrees of freedom 
of the electromagnetic field, e.g., quadrature and phase \cite{RMP,paris}.
In Ref.\ \cite{CS} it was shown that even semi-classical states as coherent states
can be employed for QKD.
Up to 2002, it was believed that QKD could not possibly work for channel 
loss above $3$ dB. This beliefs was proven wrong in Ref.\ \cite{Norbert}.
Indeed, if it is Bob to send error correcting information to Alice 
(reverse reconciliation \cite{Maurer}) then one can in principle obtain
secrecy in the presence of arbitrary high loss \cite{3dB,3dBNature,EBe,RCI}.
In 2006 it was shown that switching between two different bases for 
state preparation and measurement is not necessary for CV QKD protocols 
based on coherent state preparation and heterodyne detection \cite{Chris}.
Thus with no-switching protocols one can avoid to sacrifice part of the data 
during the sifting phase.

Only very recently, MDI QKD 
has been introduced as a framework to prevent
side-channel attacks on the measurement devices \cite{Scarani,Makarov}.
In fact, in MDI QKD the honest users are only required to prepare
quantum states, but not to measure them, as the measurement is delegated 
to an untrusted relay \cite{Curty,Sam} \cite{NOTA38}.  
In this way one does not need to make any assumption on the
measurement device: a way to guarantee security against side-channel
attacks.

\begin{figure}[ptb]
\centering
\includegraphics[width=0.45\textwidth]{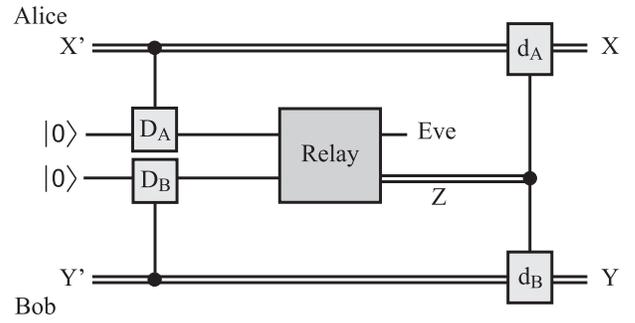}
\caption{The figure shows the scheme of the CV MDI QKD protocol of Ref.\ \cite{carlo}.
Single lines represent bosonic modes, double lines classical variables.
Time flows from left to right.
Alice and Bob initially prepare coherent states by applying displacement
operators $D_A$, $D_B$ to the vacuum state $|0\rangle$, according to the value 
of their local classical variables.
The coherent states are collected by the relay that, through some (in principle unknown) physical
transformation, outputs a classical variable $Z$ and gives to Eve quantum
side information. Finally, Alice and Bob apply {\it classical} displacement
$d_A$, $d_B$, conditioned on the value of $Z$, 
to their local classical variables.} \label{MDI-PM}
\end{figure}

{\it Description of the CV MDI QKD protocol:--}
CV MDI QKD plays a central role to show that in CV QKD 
all the raw data can be used for both parameter estimation and 
secret key generation.
Therefore, before proceedings, we need to recall the details of 
the CV MDI QKD protocol put forward in Ref.\ \cite{carlo}.
The security of this protocol was proven in Ref.\ \cite{carlo} in
the asymptotic limit, and in Ref.\ \cite{compo} in a finite-size,
composable setting.
The protocol, schematically summarized in Fig.\ \ref{MDI-PM}, develops in five steps:
\begin{enumerate}

\item
{\it Coherent states preparation.} 
Alice and Bob locally prepare $2 n$ coherent states,
with complex amplitudes denoted as 
$\alpha' = ( q'_A + i p'_A )/2$ and $\beta' = ( q'_B + i p'_B )/2$ \cite{NOTAvacuum}.
The local variables $X' \equiv (q'_A, p'_A)$
and $Y' \equiv (q'_B, p'_B)$
are drawn i.i.d.\ from zero-mean, circular symmetric, 
Gaussian distributions with variances $V_A$ and $V_B$, respectively.

\item 
{\it Operations of the relay.}
The $2n$ coherent states are sent to a central relay. 
For each pair of coherent states received the relay 
publicly announces a variable $Z$ with complex value 
$\gamma = ( q_Z + i p_Z )/2$.
If the relay is trustworthy, it operates a (lossy and noisy) 
CV Bell detection \cite{Vaidman,SamKimble,review}.

\item
{\it Parameter estimation.}
Alice and Bob estimate the covariance matrix (CM) of the variables
$( q'_A , p'_A , q'_B, p'_B, q_Z, p_Z )$.
We remark that the property of extremality of Gaussian states implies
that the knowledge of the CM is sufficient to assess the security of 
the protocol \cite{Wolf,Raul}.

\item
{\it Conditional displacements.}
Alice and Bob define the displaced variables
$X=(q_A,p_A )$ and $Y=( q_B ,p_B )$ as follows:
\begin{align}
q_A & = q'_A - g_{q'_A}(\gamma)  \, , \, \, p_A = p'_A - g_{p'_A}(\gamma)  \, , \label{dsqA} \\
q_B & = q'_B - g_{q'_B}(\gamma)  \, , \, \, p_B = p'_B - g_{p'_B}(\gamma)  \, , \label{dspB} 
\end{align}
where $g_\star$, for $\star = q'_A, p'_A, q'_B, p'_B$,
is an affine function of $\gamma$.
The variables $X$, $Y$ represent the
local raw keys of Alice and Bob, respectively.

\item
{\it Classical post-processing.}
To conclude the protocol, the raw keys 
are post-processed for error correction and privacy amplification. 

\end{enumerate}

As a matter of fact, we have defined not just one protocol, but a whole 
family of CV MDI QKD protocol: one for each choice of the affine functions $g_\star$'s.
In particular, the CV MDI protocol of Ref.\ \cite{carlo} is defined for 
an optimal choice of the functions $g_\star$
(which for completeness is derived below).


{\it Parameter estimation with almost no public communication:--}
The CV MDI QKD protocol described above has two main characteristic features.
The first is that Alice and Bob do not apply any measurement:
the only measurement is performed by the relay, which is assumed to be untrusted.
This property defines the protocol as MDI, as we are not making
any assumption on the measurement actually performed by the relay.
The second feature represents the main contribution of this Letter:
the estimation of the CM of 
$( q'_A , p'_A , q'_B, p'_B, q_Z, p_Z )$ can be done locally by either
Alice or Bob. 
Obviously, Alice and Bob know, by definition of the protocol,
the variances of $q'_A$, $p'_A$, $q'_B$, $p'_B$.
Also, Alice can locally estimate the correlation terms
$\langle q'_A q_Z \rangle$, $\langle q'_A p_Z \rangle$,
$\langle p'_A q_Z \rangle$, $\langle p'_A p_Z \rangle$, from her
local data and from the amplitude $\gamma = (q_Z + i p_Z)/2$
that have been publicly announced by the relay \cite{NOTApe}.
Similarly, Bob can locally estimate
$\langle q'_B q_Z \rangle$, $\langle q'_B p_Z \rangle$,
$\langle p'_B q_Z \rangle$, and $\langle p'_B p_Z \rangle$.
This implies that all the entries of the CM of 
$( q'_A , p'_A , q'_B, p'_B, q_Z, p_Z )$ can be locally
estimate by either Alice or Bob, without the need of
public communication.

We remark that here we do not need to specify the explicit
procedure to obtain the confidence intervals for the estimated
parameters. 
This can be done in many different ways. 
For example, under the additional assumption that the variables
$( q'_A , p'_A , q'_B, p'_B, q_Z, p_Z )$ are Gaussian,
one can proceed as described in Refs.\ \cite{compo,PP,Usenko}.
Otherwise, one can apply the statistical analysis of
Ref.\ \cite{Lev1} which does not assume Gaussianity. 
In either case the required data for the estimation of the
correlation terms are all locally available to the users.

Finally, the CM of $( q_A, p_A, q_B, p_B)$ can be computed directly
from the CM of $( q'_A , p'_A , q'_B, p'_B, q_Z, p_Z )$ by exploiting the
relations (\ref{dsqA})-(\ref{dspB}).
In conclusion, the CM of $( q_A, p_A, q_B, p_B)$ can be estimated 
only exploiting locally available information since, as we show in
the following section, the functions $g_\star$ can be also computed 
from local data only.
This is ultimately possible because in an MDI QKD the correlations 
between Alice's and Bob's raw keys are post-selected by the relay. 
Therefore, the public variable $Z$ contains all the information about the
correlations between Alice and Bob and is thus sufficient, together
with the local data, to estimate the CM.


{\it Optimal conditional displacements:--}
For completeness we now derive the optimal choice for the displacement
functions $g_\star$ \cite{NOTAsimple}.
At the parameter estimation stage, Alice and Bob locally estimate
the CM of $(q'_A, p'_A, q'_B, p'_B, q_Z, p_Z)$:
\begin{align}\label{VABZ}
V_{A'B'Z} = \left(
\begin{array}{ccc}
V_A \mathbf{I}              & 0                          & \mathbf{c}_{AZ} \\
0                          & V_B \mathbf{I}              & \mathbf{c}_{BZ}  \\
\mathbf{c}_{AZ}^\mathsf{T} & \mathbf{c}_{BZ}^\mathsf{T} & \boldsymbol{v}_Z
\end{array}
\right) \, ,
\end{align}
where $\mathbf{I}$ denotes the two-dimensional identity matrix, 
\begin{align}
\boldsymbol{v}_Z = \left(
\begin{array}{cc}
\langle q_Z^2 \rangle & \langle q_Z p_Z \rangle \\
\langle q_Z p_Z \rangle & \langle p_Z^2\rangle
\end{array}
\right)
\end{align}
is the empirical CM of $(q_Z, p_Z)$, and
\begin{align} 
\mathbf{c}_{AZ} = \left(
\begin{array}{cc}
\langle q'_A q_Z \rangle & \langle q'_A p_Z \rangle \\
\langle p'_A q_Z \rangle & \langle p'_A p_Z\rangle
\end{array}
\right) , 
\mathbf{c}_{BZ} = \left(
\begin{array}{cc}
\langle q'_B q_Z \rangle & \langle q'_B p_Z \rangle \\
\langle p'_B q_Z \rangle & \langle p'_B p_Z\rangle
\end{array}
\right)
\end{align}
are the correlation terms.

We remark that the variables $(q'_A,p'_A,q'_B,p'_B)$ are uncorrelated 
with known variances $V_A$, $V_B$ by definition of the protocol, while all
the entries involving the publicly known variables $(q_Z, p_Z)$ 
must be estimated from the data.


The optimal choice for the displacements in Eqs.\ (\ref{dsqA})-(\ref{dspB})
is the one that minimizes the correlations between Alice's and Bob's variables and
$\gamma = ( q_Z + i p_Z)/2$. 
Therefore we put, for $\star = q'_A, p'_A, q'_B, p'_B$,
\begin{equation}
g_\star(\gamma) = u_\star \, q_Z + v_\star \, p_Z \, ,
\end{equation}
and require that $u_\star$ and $v_\star$ are chosen in such a way that
\begin{align}
\langle q_Z q_A \rangle = \langle p_Z q_A \rangle = \langle q_Z p_A \rangle = \langle p_Z p_A \rangle & = 0 \, ,\\
\langle q_Z q_B \rangle = \langle p_Z q_B \rangle = \langle q_Z p_B \rangle = \langle p_Z p_B \rangle & = 0 \, ,
\end{align}
which implies
\begin{align}
\langle \star \, q_Z \rangle & = u_\star \, \langle q_Z^2 \rangle + v_\star \langle q_Z p_Z \rangle  \, , \\
\langle \star \, p_Z \rangle & = u_\star \, \langle q_Z p_Z \rangle + v_\star \langle p_Z^2 \rangle  \, .
\end{align}
Solving for $u_\star$ and $v_\star$ we obtain
\begin{align}
u_\star & = \frac{ \langle \star \, q_Z \rangle \langle p_Z^2 \rangle - \langle \star \, p_Z \rangle \langle q_Z p_Z \rangle }{ \langle p_Z^2 \rangle \langle q_Z^2\rangle - \langle q_Z p_Z \rangle^2}\, , \label{u}\\
v_\star & = \frac{ \langle \star \, p_Z \rangle \langle q_Z^2\rangle - \langle \star \, q_Z \rangle \langle q_Z p_Z \rangle }{ \langle q_Z^2 \rangle \langle p_Z^2 \rangle - \langle q_Z p_Z \rangle^2  } \, . \label{v}
\end{align}
With this choice of the parameters $u_\star$, $v_\star$ the displaced variables
$(q_A,p_A,q_B,p_B)$ are independent of $(q_Z,p_Z)$.
We remark that in this way the 
CM $V_{AB}$ of $(q_A, p_A, q_B, p_B)$ equals the  
conditional CM of $(q'_A, p'_A, q'_B, p'_B)$ conditioned 
on $(q_Z, p_Z)$ (see Ref.\ \cite{carlo}).


As an example, put $V_A = V_B = 2N$ and suppose that the relay applies a Gaussian transformation
that consists of (see Ref.\ \cite{carlo}): first attenuating the signals from Alice and Bob by
an attenuation factor $\eta$;
and then perform an ideal, noiseless, CV Bell detection.
In this case one obtains:
\begin{equation}
-u_{q'_A} = v_{p'_A} = u_{q'_B} = v_{p'_B} = \frac{N}{\eta N + 1/2} \sqrt{ \frac{\eta}{2} } \, . 
\end{equation}
Other numerical examples are discussed in Ref.\ \cite{compo}.

\begin{figure}[ptb]
\centering
\includegraphics[width=0.45\textwidth]{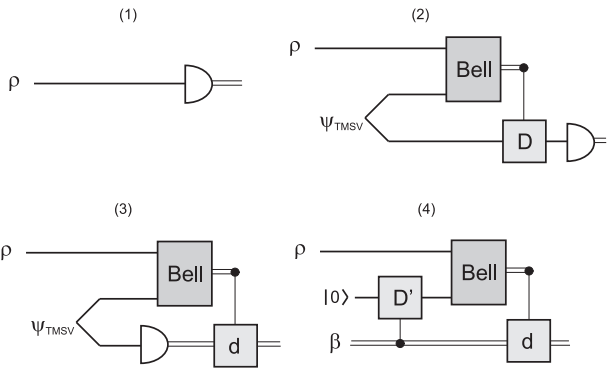}
\caption{The figures show: 
(1) direct heterodyne detection; and
(4) MDI-inspired detection, obtained when the relay is given to the receiver Bob.
Single lines indicate bosonic modes, double line classical variables.
(2) and (3) show intermediate configurations that we exploit to 
prove the equivalence, up to an arbitrarily small error, between (1) and (4).
Notice that in (4) we have described the preparation of a coherent state $|\beta\rangle$
of amplitude $\beta$ as the
application of a displacement $D'$ on the vacuum, where the amplitude of the
displacement is determined by a classical variable $\beta$.
} \label{EBequiv}
\end{figure}

{\it From MDI to one-way CV QKD:--}
In the MDI framework, Alice and Bob send quantum states to a
central relay, which is untrusted and possibly operated by an
eavesdropper.
On the other hand, in a one-way QKD protocol, Alice sends a 
quantum state $\rho$ to the receiver Bob, who measures it, 
typically by homodyne or heterodyne detection, as shown in Fig.\ \ref{EBequiv}(1).

First of all, an MDI protocol can simulate with arbitrary
high precision any one-way protocol. In fact, if the relay
is given to Bob, he can use it to teleport the signals 
from Alice into his lab, as shown in Fig.\ \ref{EBequiv}(2).
Clearly, ideal CV teleportation requires Bob to employ
as teleportation resource a two-mode squeezed vacuum (TMSV) 
state $\psi_\mathrm{TMSV}$ with infinite squeezing \cite{Vaidman,SamKimble,review} .
Otherwise, for any finitely squeezed TMSV state, the scheme in Fig.\ \ref{EBequiv}(2) 
simulates that in Fig.\ \ref{EBequiv}(1) with up to additive Gaussian noise \cite{Ban,Andy,Scroto,Qmetro,fr}.
Since the displacement operation commutes with heterodyne detection,
to apply a displacement $D$ and then measure by heterodyne detection [as in Fig.\ \ref{EBequiv}(2)] 
is equivalent to first measure and then displace the (classical) outcome of the
measurement [as in Fig.\ \ref{EBequiv}(3)].
Finally, it is well known that measuring by heterodyne detection one mode of an entangled
pair in a TMSV state, conditionally prepares the other mode in a coherent state \cite{EBe},
this implies the equivalence between the schemes in Fig.\ \ref{EBequiv}(3) and Fig.\ \ref{EBequiv}(4).
In conclusion, the MDI protocol in Fig.\ \ref{EBequiv}(4) can simulate the
one-way CV QKD. 
If the complex amplitude $\beta$ is sampled from a Gaussian distribution 
with finite variance $V_B$, then one simulates a noisy version of the QKD protocol,
whereas the noiseless case is obtained in the limit that $V_B \to \infty$.

{\it Discussion:--}
As shown above, parameter estimation in CV MDI QKD can be
performed with almost no public communication
because correlations are post-selected by the central relay.
This condition is necessary but would not be sufficient without
the additional property that in CV QKD the knowledge of the
CM of the quadratures is sufficient to asses the security of the
protocol.
In particular, the conditional 
probability distribution $P(XY|Z)$, which is the relevant quantity
for assessing the security of the protocol \cite{Sam}, can be estimated
from the elements of the CM alone. 
In other words, the knowledge of the marginal probability distributions
$P(XZ), P(YZ)$ is sufficient to know $P(XYZ)$. 
This is the property that we have exploited above.

It is meaningful to ask whether one can perform parameter estimation 
without public communication also in the case of discrete-variable
MDI QKD. The answer to this question is negative because, although correlations are 
still post-selected by the relay, the knowledge of the marginals is no longer
sufficient to characterize the protocol.
Consider for example the qubit MDI protocol of Ref.\ \cite{Curty}, which can 
be viewed as an MDI version of BB84, where the variables $X$ and $Y$
assume values in $\{ 0,1 \}$, and $Z \in \{ 0, 1, 2, 3 \}$ is the output
of qubit Bell detection.
One can easily check that in this setting the marginal
probability distributions $P(XZ), P(YZ)$ do not uniquely determine
$P(XYZ)$. 


{\it Conclusions:--}
The list of conceptual breakthroughs in the history of QKD includes the discoveries that 
reverse reconciliation allowed to beat the $3$dB barrier, that coherent states
were suitable for QKD despite being semiclassical, and that CV QKD did not require
switching between different bases for encoding and measurement, 
thus allowing us to skip the sifting phase.

This Letter presents one new conceptual development of CV QKD, namely that
the whole raw keys can be used for both parameter estimation and secret key extraction.
This finding removes the tradeoff between secret key rate and
accuracy of the parameter estimation in the finite-size regime of QKD. 
Unlike other works \cite{Pacher,Lev1}, here we do not need an initial rough
estimate of the signal-to-noise ratio nor we require 
to perform error correction before parameter estimation.

Such a property is first obtained for CV MDI QKD protocols as a consequence of 
the fact that correlations between Alice and Bob are encoded in the variable 
that is publicly announced by the relay --- even though such a variable does not 
contain information about the secret key.
Since CV MDI QKD can simulate one-way CV QKD protocols
with arbitrary precision,
it then follows that the whole raw key can be used for 
both parameter estimation and secret key generation for
this class of CV protocols as well.

\acknowledgments

This work was supported 
by the 
Innovation Fund Denmark 
within the
Quantum Innovation Center Qubiz,
and by the
UK Quantum Communications hub (EP/M013472/1).
C.L. acknowledges the scientific support received from the 
Quantum Physics and Information Technology Group (QPIT).


\begin{thebibliography}{9}
\bibitem{NOTAvacuum}
We put $\hbar = 1$ and assume the commutation relations of the form
$[ q , p ] = 2i$ \cite{paris}. In this way the output of homodyne 
detection over the vacuum state is a Gaussian variable with variance $1$.
\bibitem{NOTApe}
To make this idea more concrete, consider for example the
task of estimating $\langle q'_A q_Z \rangle$ from Alice's local
empirical data ${q'_A}_j$'s and the data publicly broadcast by the relay
${q_Z}_j$'s. The estimator for $\langle q'_A q_Z \rangle$ is then
$ \frac{1}{n} \sum_{j=1}^n {q'_A}_j {q_Z}_j$, which can be
locally computed by Alice.
\bibitem{NOTAsimple}
To simplify the notation, we assume that $\langle q_Z \rangle = 
\langle p_Z \rangle = 0$ (one can always redefine the variables $q_Z$, $p_Z$
to ensure this property is verified).
\end{thebibliography}
\end{document}